%% file: main.tex
\documentclass{article}
\usepackage{spconf,amsmath,epsfig}

\providecommand{\doi}[1]{doi: {\footnotesize \href{http://dx.doi.org/#1}{\path{#1}}}}

\usepackage[pdftex=true,breaklinks=true,hidelinks=true,colorlinks=true,citecolor=blue]{hyperref}

\usepackage[square,numbers]{natbib}

\makeatletter
\renewcommand\NAT@bibsetnum[1]{\settowidth\labelwidth{\@biblabel{#1}}%
  \setlength{\leftmargin}{\bibindent}\addtolength{\leftmargin}{\dimexpr\labelwidth+\labelsep\relax}%
  \setlength{\itemindent}{-\bibindent}%
  \setlength{\listparindent}{\itemindent}
  \setlength{\itemsep}{\bibsep}\setlength{\parsep}{\z@}%
  \ifNAT@openbib
  \addtolength{\leftmargin}{\bibindent}%
  \setlength{\itemindent}{-\bibindent}%
  \setlength{\listparindent}{\itemindent}%
  \setlength{\parsep}{0pt}%
  \fi
}
\makeatother

\usepackage[utf8]{inputenc} 
\usepackage[T1]{fontenc}    
\usepackage{hyperref}       
\usepackage{url}            
\usepackage{array,booktabs}       
\usepackage{amsfonts,amsmath}       
\usepackage{nicefrac}       
\usepackage{microtype}      
\usepackage{mathtools}

\usepackage[dvipsnames,svgnames]{xcolor}
\usepackage[normalem]{ulem}
\newif{\ifhidecomments}
\usepackage{tumcolors}

\usepackage{tikz,pgf}
\usetikzlibrary{calc}
\usetikzlibrary{fit}
\usetikzlibrary{positioning}
\usepackage{pgfmath}

\usepackage{pgfplotstable}

\usepackage{svg}

\input{images/pgfstyles.pgf}

\usepackage{subcaption}
\def\x{{\mathbf x}}

\newcommand{\eg}{\textit{e{.}g{.}}, }
\newcommand{\cf}{\textit{cf{.}}\,}

\usepackage[
  capitalize,
]{cleveref}
\usepackage{csquotes}
\usepackage{enumitem}

\usepackage[
colorinlistoftodos,
color={red!100!green!33},
shadow,
]{todonotes}
\presetkeys{todonotes}{%
	size=\tiny,
}{}
\usepackage{setspace}
\usepackage{xargs}
\newcounter{mycomment}
\newcommandx{\note}[3][
1=, 
3=, 
]{%
	\refstepcounter{mycomment}{%
		\setstretch{0.7}
		\todo[#3]{\textbf{#1\themycomment:} #2}%
	}%
}

\newcommand{\reduline}{\bgroup {\textcolor{red}\ULdepth=-.55ex} \ULset}

\usepackage{fancyvrb}

\title{EUROCROPS: A PAN-EUROPEAN DATASET FOR TIME SERIES CROP TYPE CLASSIFICATION}
\name{%
  Maja Schneider\textsuperscript{1}%
  \thanks{%
    M. Schneider was funded by the Federal Ministry for Economic Affairs and Energy
    following a resolution of the German Bundestag
    under reference 50EE1908.%
  },
	Amelie Broszeit\textsuperscript{2},
  Marco Körner\textsuperscript{1}
}
\address{School\\
	Department\\
	Address}
\address{%
  \parbox[t]{.55\linewidth}{\centering%
    \textsuperscript{1}~%
    Chair of Remote Sensing Technology\\
    Department of Aerospace and Geodesy\\
    Technical University of Munich (TUM),
    Munich, Germany
  }
  \parbox[t]{.35\linewidth}{\centering%
    \textsuperscript{2}~%
    GAF AG\\
    Munich, Germany
  }
}
\input{figures.tex}

\input{tables.tex}
\input{notation.tex}

\begin{document}
%
\maketitle
\begin{abstract}
We present \eurocrops, a dataset based on self-declared field annotations for training and evaluating methods for crop type classification and mapping, together with its process of acquisition and harmonisation.
By this, we aim to enrich the research efforts and discussion for data-driven land cover classification via Earth observation and remote sensing.
Additionally, through inclusion of self-declarations gathered in the scope of subsidy control from all countries of the European Union (EU), this dataset highlights the difficulties and pitfalls one comes across when operating on a transnational level.
We, therefore, also introduce a new taxonomy scheme, \brand{HCAT-ID}, that aspires to capture all the aspects of reference data originating from administrative and agency databases.
To address researchers from both the remote sensing and the computer vision and machine learning communities, we publish the dataset in different formats and processing levels.
\end{abstract}
\begin{keywords}
big data, crops, analysis ready data, machine learning, Earth observation
\end{keywords}
\section{INTRODUCTION}
\label{sec:intro}

\begin{figure}[t]
  \centering
  \includegraphics[width=0.8\columnwidth]{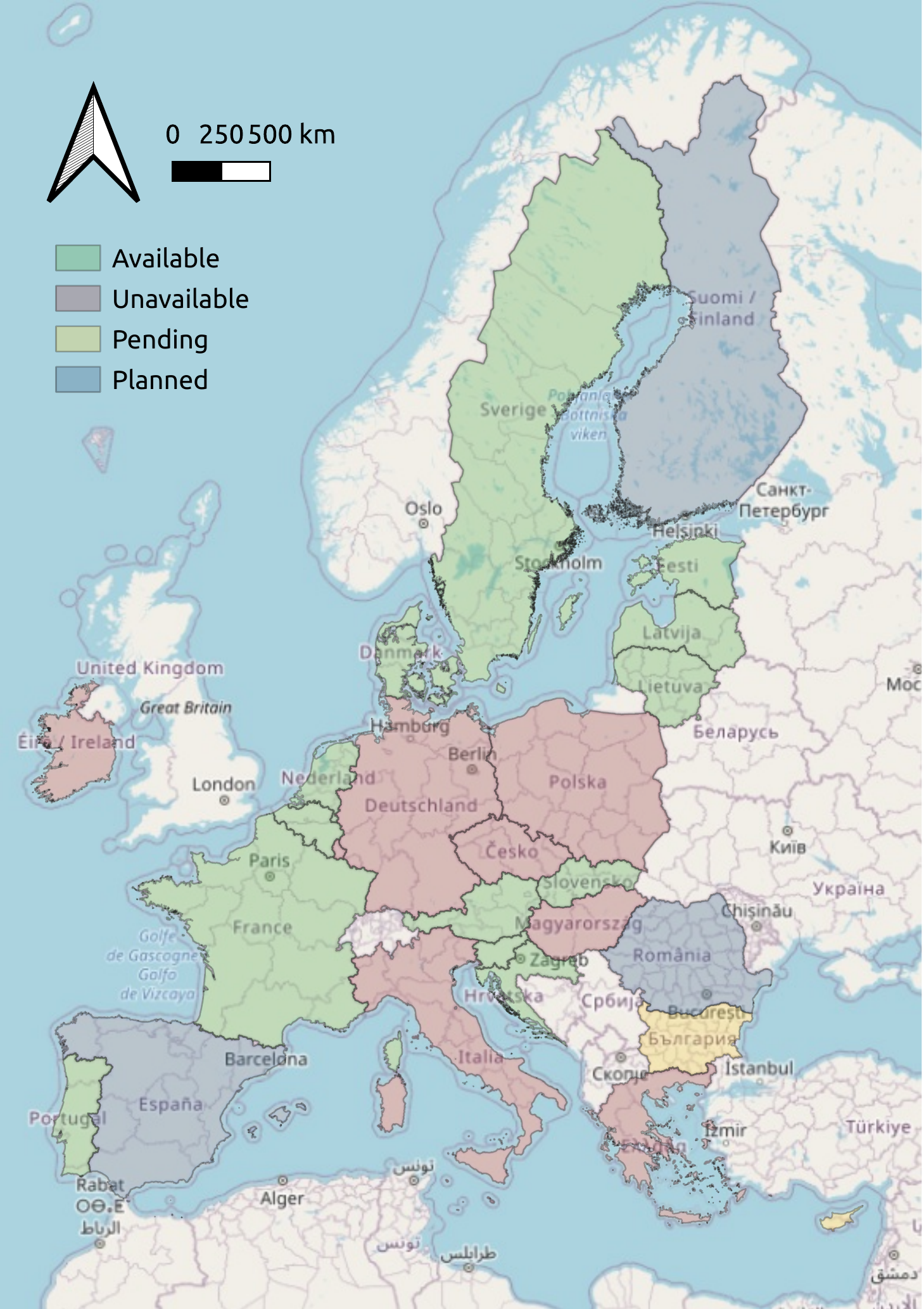}
  \caption{13 of all 27 European Union member states released their agricultural parcel information reference data by the end of 2020 for research.
    A constantly updated and interactive version of this map can be found on \href{www.eurocrops.tum.de}{www.eurocrops.tum.de}.}
  \label{fig:countries}
\end{figure}

Exploration of data-driven models in the domain of Earth observation currently engages an entire branch of scientists, encouraging them to apply insights and developments from recent machine learning research to tackle problems prevalent in remote sensing.
However, continued development in the area of satellite data analysis, notably using learning-based techniques, suffers from significant restraint due to insufficient diversity, quantity, and reliability of annotated reference data.
In conjunction with the peculiarities of Earth observation data, this creates a high barrier to entry and consequently deters those from outside the field from contributing their ideas to this domain.
With \eurocrops, we strive to tackle these issues for the application of crop type classification, particularly by using reliable self-declarations and not retrospective handmade annotations.

\section{EUROCROPS}
\label{sec:ec_approach}

With the development of \eurocrops, we aim to build up a harmonised reference dataset containing agricultural parcel information data together with the crop type on each parcel, based on farmers' self-declarations following the European subsidy control.
It includes all European Union member states that make this data publicly available for research purposes.
We want to broaden the target audience of this dataset beyond those interested in remote sensing and agriculture, intending to address computer scientists and data analysts without a strong background in Earth observation and remote sensing.
Thus, \eurocrops provides its data in analysis-ready formats.

\subsection{Data Acquisition}
\label{ssec:data_acqu}

To achieve pan-European coverage with reference data, we conducted intensive enquiries about all 27 EU member states and their position regarding the data.
As most of the countries, at that time, did not offer direct ways to download files in shapefile or \geojson format, we got in touch with ministries, agricultural departments, and authorities from 24 states.

\subsection{Participating Countries}
\label{ssec:part_countries}

\tabAvailability

By April 2021, it was possible to obtain reference data from 13 countries of the European Union, while another three are planning to release the data within 2021.
\Cref{fig:countries} provides a map of all participating countries, while the corresponding detailed line-up is listed in \cref{tab:countries}.
The acquisition process will be continued in the next years to
not only achieve maximum regional but also temporal coverage.

\subsection{Taxonomy}
\label{ssec:taxonomy}

\figTaxonomy

Although the European Union aims to achieve a common agricultural policy, current reference data hardly comes in a harmonised and uniform format, or even with English crop names.
One of the biggest challenges of creating this dataset was to find one foundation of taxonomy into which all country-dependent schemes fit into.
Consequently, we adapted the \brand{EAGLE} matrix \cite{arnold2013eagle} developed by the European Environment Agency.
Its \emph{Characteristics} (CH) block represents additional land cover classes, including a section \emph{Crop Type} with a segregation into classes according to EU regulation (EC) 1200/2009, annex II, chapter 2 \cite{ec2009eu}.
To ensure optimal granularity, supplementary classes were added to the matrix.
In addition, we propose an eight-digit taxonomy system,
the \emph{Hierarchical Crop and Agriculture Taxonomy Identifier (HCAT-ID)},
in which the classes are represented numerically, while also encoding the hierarchical precision of the chosen crop names.
It ranges, \eg from
\class{Cereals} as \HCAT{3}{3}[1][01][00][0] (level 4)
to
\class{Summer Oats} as \HCAT{3}{3}[1][01][05][1] (level 6),
or generally
\begin{equation*}
	\underbracket{[3][3]}_{\substack{\text{position in}\\ \text{\brand{EAGLE} matrix}}}\text{--}
	\underbracket{\text{X}}_\text{Level 3}\text{--}
	\underbracket{\text{XX}}_\text{Level 4}\text{--}
	\underbracket{\text{XX}}_\text{Level 5}\text{--}
	\underbracket{\text{X}}_\text{Level 6}.
\end{equation*}
This way, it is easily possible to cut the \brand{HCAT-ID} at the desired level of granularity on-the-fly and to use the label without any preprocessing and further refinement.
By eliminating the last digit, for instance, the seasonal attribute of the crops can be omitted and a differentiation will only be made between \class{Rye} and \class{Oats}.
The distinction is also shown in \cref{fig:tax}, where each concentric circle represents one of the levels.
They consist of blocks that are either a crop group, which include several classes from an outward circle, or an atomic crop class itself.
For further information, the entire taxonomy and any updates on it, will also be published on our website and \github.

\section{DEMO DATASET AND ANALYSIS}
\label{sec:ec_demo_dataset}

For a first study, a subset of the available data was gathered, harmonised and the corresponding \sentinel images downloaded.
By having a meaningful foundation and already being aware of all the difficulties, extending the dataset to the entire region of Europe will then follow a designated route.

\subsection{Regions}
\label{ssec:regions}

\figTestRegions
\tabCountryTiles

The regions for the demo dataset were chosen such that the agricultural diversity of Europe and its influence on the outcome of classification methods can be evaluated:
Firstly, by choosing the adjacent countries \emph{Austria} and \emph{Slovenia}, the impact of national borders and, therefore, expected differences in cultivation of agricultural land is addressed.
These two countries share approximately the same climate conditions, leading to the choice of \emph{Denmark} as a third remote region for the demonstration.
This way, it is possible to examine the effect of the latitude on the accuracy of crop type classification algorithms.
While the declarations from \emph{Austria} and \emph{Slovenia} were surveyed for the year 2020, \emph{Denmark} only released the data from 2019.
In order to keep the satellite data processing straightforward and use roughly an equal areal coverage of each region, one \sentinel tile per country was chosen, as visualised in \cref{fig:testregions} and listed in \cref{tab:sentineltilemnames}.
The resulting dataset includes nearly one million field parcels and their annotations.

\subsection{Class Distributions}
\label{ssec:class_dist}

\newsavebox{\imagebox}%
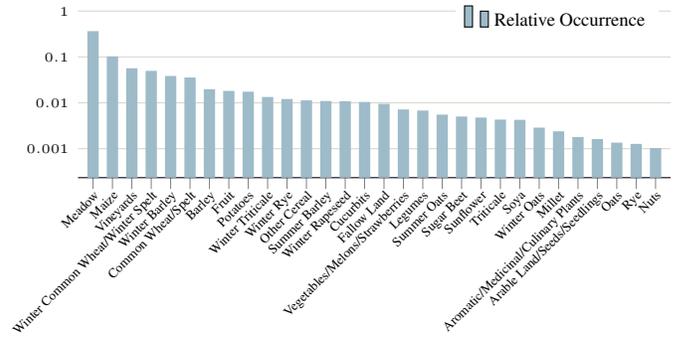
\begin{figure}
  \centering
    \centering\raisebox{\dimexpr\ht\imagebox-\height}{%
      \input{images/dataset_statistics_slovenia.pgf}%
    }%


  \caption{%
    The strong imbalance of the top-30 crop classes is indicated by an almost linear descent of relative occurrences while using a $\log$-scale. This property adds an additional challenge to the dataset, as most data-driven methods tend to rely on a balanced training input.
  }
  \label{fig:dataset_stats}
\end{figure}

\figPies

Similarly to other crop type datasets covering the temperate climate zone, an evident class imbalance towards \class{Meadow} is present in our demo dataset as well.
\Cref{fig:dataset_stats} illustrates this trend by showing a near-linear decrease of crop class occurrences in logarithmic scale.
In addition, the different geographic regions, as well as the impact of national borders, give rise to different class distributions amongst the demo areas.
As mentioned before, harmonising the data also leads to discrepancies between the occurring classes.
\emph{Denmark} and \emph{Slovenia}, for instance, are both only utilising the level 5 crop class \class{Common Wheat/Spelt}, while \emph{Austria} includes seasonal level 6 cultivation attributes as \class{Winter}, shown in the diagrams for classes $4$ and $6$ in \cref{fig:pies}.
Of all available \brand{HCAT} groups, our harmonised demo dataset uses 43 classes in the granularity reported by the respective authorities of the countries.

\subsection{Train-Test Split}
\label{ssec:train_test}

Earth observation satellite imagery is likely to be heavily influenced by spatial auto-correlation, which makes it a priority to deliberate on the choice of training and test split as early as possible.
This implies to refrain from using adjacent parcels when training and testing a machine learning algorithm for addressing remote sensing problems.
Therefore, we extracted one test area from \emph{Austria} and another one from \emph{Denmark}, following the \emph{Nomenclature of Territorial Units for Statistics (NUTS)} scheme.
One additional region in \emph{Austria} has been excluded for future benchmarking purposes and challenges.
Visually, this dissociation is shown in \cref{fig:testregions}.

\subsection{Remote Sensing Dataset}
\label{ssec:rs_dataset}

To address the remote sensing community, we publish our harmonised reference vector data in \geojson format including the geo-referenced geometry of the parcels and the corresponding crop cultivated on this parcel.
This way, any available satellite data can be used together with the labels.
Additionally, corresponding \sentinel imagery is available.

\subsection{Time Series Dataset}
\label{ssec:ts_dataset}

  \begin{figure}
    \centering
    \resizebox{.75\columnwidth}{!}{\input{images/parcels.pgf}}
    \caption{
      From each parcel, one representative pixel is chosen and all its $C=13$ bands are examined over $T$ timesteps.
      This structure is embedded into a \filename{HDF5} file, where the rows and columns correspond to the parcels and the timesteps respectively.
      In each cell, the 13 raw reflectance values for that representative pixel are stored.
    }
    \label{fig:data_struct}
  \end{figure}
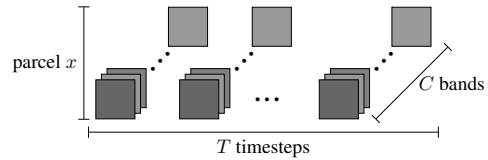

As we also wanted to create a dataset to challenge the entire machine learning community, and notably those unfamiliar with remote sensing and Earth observation satellite data, a second, much more compact dataset will be released.
In doing so, we got rid of the spatial component and concentrated on an easy-to-use and analysis-ready data format including two types of files: \filename{csv} files with the crop labels and \filename{HDF5} files with a representative reflectance value from a \sentinel L1C product for each parcel and band.
These 13 values, each corresponding to one \sentinel spectral band, were then gathered for each timestep within the observation period, resulting in a dataset ready for time series analysis.
The structure of the reflectances is visualised in \cref{fig:data_struct}.
We want to point out that we purposely used \emph{raw top-of-atmosphere} satellite data, including clouds and without atmospheric corrections, as any processing induces bias to the data.
We hope to give methods the chance to explore all available underlying properties themselves.

\subsection{Publication of the Demo Dataset}
\label{ssec:publication}

There are several options available on how to obtain the dataset:
Firstly, as mentioned before, \href{www.eurocrops.tum.de}{www.eurocrops.tum.de} provides the up-to-date status of the 
project, as well as download resources and 
a link to a \github repository.
Secondly, it will be made accessible through the \brand{Sinergise} \href{www.sentinel-hub.com}{Sentinel Hub} in the scope of the \href{www.globalearthmonitor.eu}{Global Earth Monitor (GEM)} project.
Lastly, we encourage scientists doing research on this topic to reach out and we will be happy to provide them the data directly.

\section{CONCLUSION}
\label{sec:conclusion}

We presented \brand{EuroCrops}, a dataset that will help in advancing the research in the domain of crop type classification.
Inspired by the success of previously published datasets from \citeauthor{russwurm2020breizhcrops} \cite{russwurm2020breizhcrops} and \citeauthor{turkoglu2021crop} \cite{turkoglu2021crop}, we strive to curate a dataset with similar properties, but a larger regional coverage and a wider range of classes.
As a first step, we publish a demo dataset covering regions in three different countries, which will then be extended in near future to contain several entire member states of the European Union.
In addition to that, we introduced an extension to the \brand{EAGLE} taxonomy scheme, which we encourage everyone to adapt and use.
Finally, our dataset shows where the pitfalls of transnational datasets lie and therefore builds a new foundation for research targeting this area.


\setlength{\bibsep}{0pt}
\small

\bibliographystyle{hunsrtnat}
\bibliography{bib/bib.bib}

\end{document}

%% file: images/pgfstyles.pgf
\usetikzlibrary{fit,positioning}%
\tikzset{%
  node distance/.append code={
    \pgfkeyssetvalue{/tikz/node distance value}{#1}
  },
}%
\tikzset{%
  every node/.append style={
    align=center,
  },
  block/.style={%
    draw,
    rounded corners,
  },%
  input/.style={
  },
  encoder/.style={
    block,
    fill=zombie,
  },
  fc/.style={
    block,
    fill=martini,
  },
  attention/.style={
    block,
    fill=calypso,
    minimum width=3cm,
  },
  mlp/.style={
    block,
    fill=aquaisland,
  },
  operation/.style={
    circle,
    draw,
    text width=3mm,
  },
  add/.style={
    operation, inner sep=2pt,
    node contents={$+$},
  },
  mult/.style={
    operation, inner sep=2pt,
    node contents={$\cdot$},
  },
  arrow/.style={
    -latex,rounded corners,
  }
}%

\pgfplotsset{select coords between index/.style 2 args={
  x filter/.code={
    \ifnum\coordindex<#1\def\pgfmathresult{}\fi
    \ifnum\coordindex>#2\def\pgfmathresult{}\fi
  }
}}

\pgfplotsset{
  bar/.style={
    draw=none,
  },
  freq bar/.style={
    bar,
  },
  crop bar/.style={
    width=1.1\linewidth,
    ybar=0.0mm,
    bar width=1.5mm,
    axis x line*=bottom,
    axis y line*=left,
    y axis line style={draw=none},
    ymajorgrids=true,
    enlarge x limits={abs=2mm},
    enlarge y limits=true,
    major grid style={draw=tumivory},
    ytick style={draw=none},
    x tick label style={font=\tiny, rotate=45, anchor=north east, inner sep=0pt},
    y tick label style={font=\tiny},
    xlabel near ticks,
    ylabel near ticks,
    xlabel={crop type class},
    ylabel={relative occurrence},
    label style={font=\scriptsize},
    legend style={
      legend columns=-1,
      font=\scriptsize,
      draw=none,
    },
  },
}

%% file: figures.tex
\newcommand{\figTaxonomy}{%
  \begin{figure}
    \centering
    \includegraphics[width=0.8\linewidth]{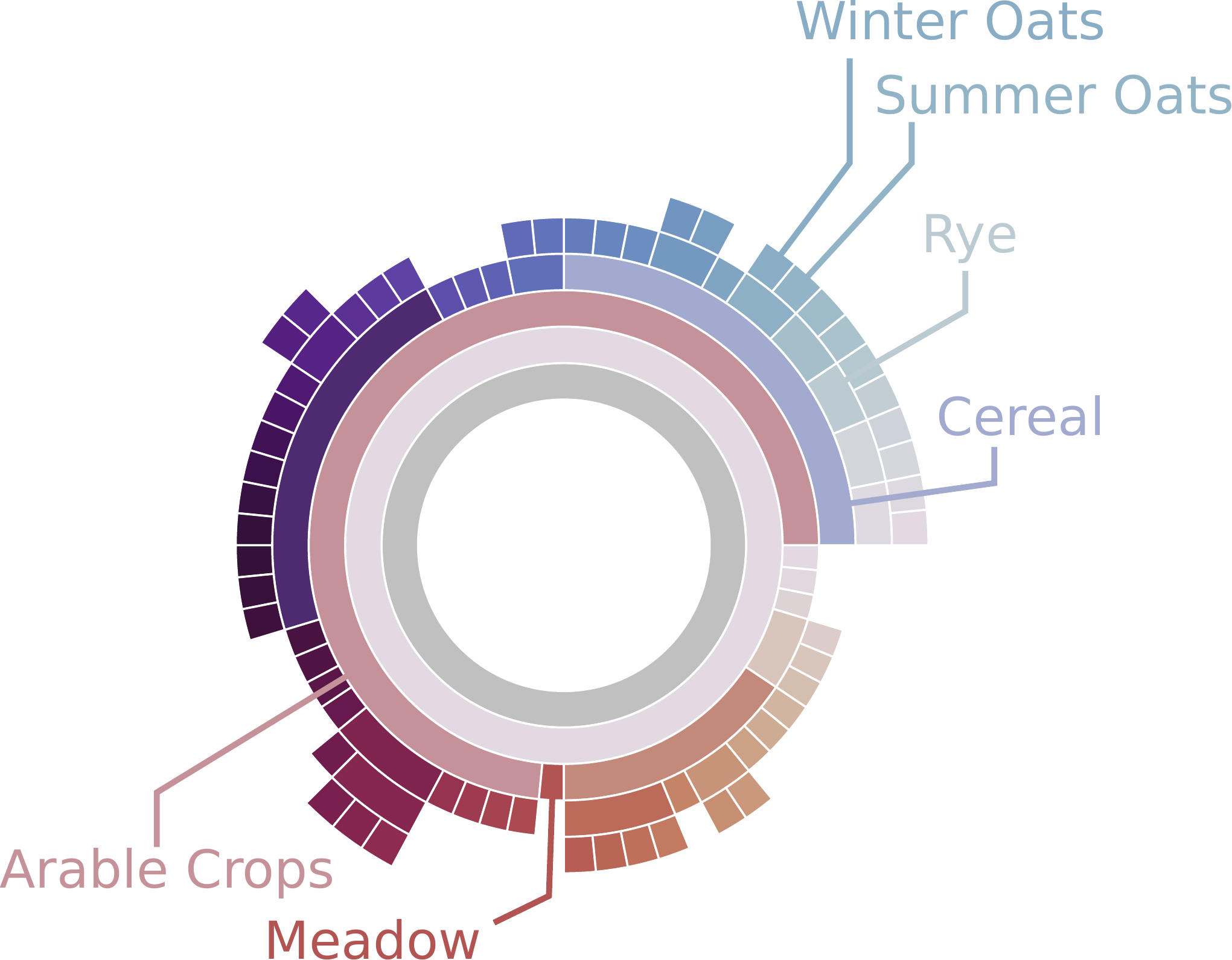}
    \caption{%
      A visual explanation of the different taxonomy levels used in the dataset.
      While the inner two circles represent the incorporation of our chosen taxonomy into the entire \mbox{\brand{EAGLE}} concept, each further level introduces a finer granularity:
      While level 3 only distinguishes between general \emph{Land Cover} types like \class{Arable Crops} and \class{Meadow}, level 6 introduces \eg seasonal changes.
      \class{Oats}, which is a level 5 class like \class{Rye}, can therefore be divided into \class{Winter Oats} and \class{Summer Oats}, as shown.
      Owing to our {HCAT-ID}, it is possible to choose the desired granularity on-the-fly, while not having to omit all details while setting up the dataset.
    }
    \label{fig:tax}
  \end{figure}
}

\newcommand{\figTestRegions}{%
\begin{figure}[t]
  \begin{minipage}[b]{.48\linewidth}
    \centering
    \includegraphics[width=\columnwidth]{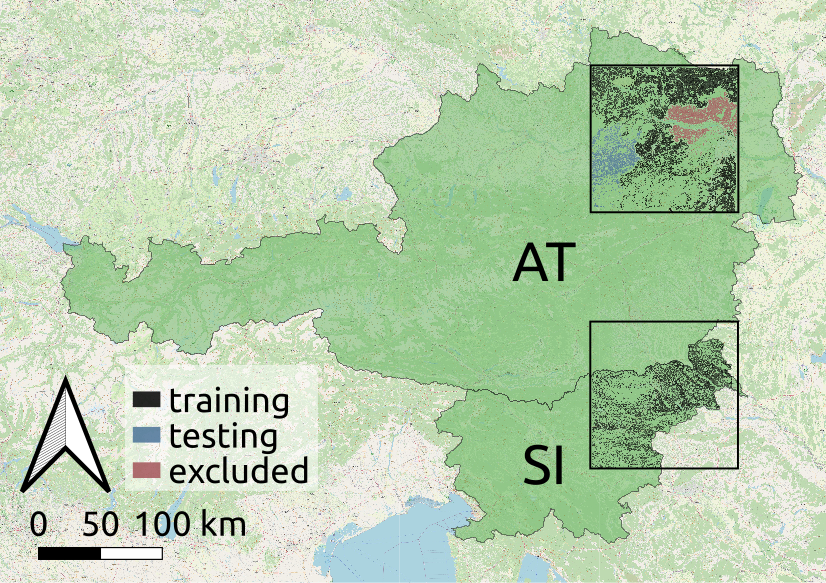}
    \subcaption{The test areas in \emph{Austria} and \emph{Slovenia} were chosen for the investigation of similarities and differences of bordering regions.\label{fig:testregions_AS}}
  \end{minipage}
  \hfill
  \begin{minipage}[b]{0.48\linewidth}
    \centering
    \includegraphics[width=\columnwidth]{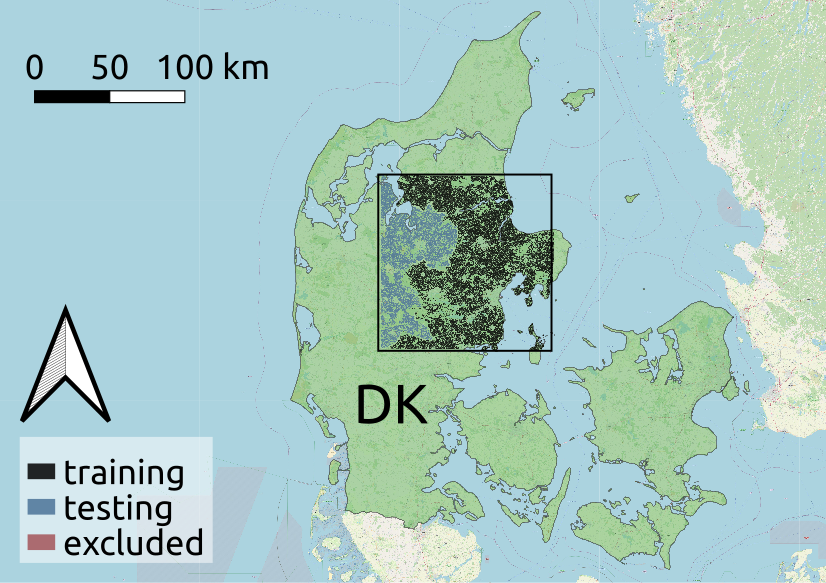}
    \subcaption{By choosing \emph{Denmark} as a third demo country, the impact of different climate conditions in one dataset can also be examined.\label{fig:testregions_D}}
  \end{minipage}
  \caption{%
    The selected test areas cover regions \subref{fig:testregions_AS} in \emph{Austria} and \emph{Slovenia}, and \subref{fig:testregions_D} in \textit{Denmark}, each corresponding to one \sentinel tile.
    In order to counteract the influence of spatial correlation between adjacent crop field parcels, the test areas were chosen in advance and regionally separated from the training areas.
  }
  \label{fig:testregions}
\end{figure}
}

\newcommand{\figPies}{%
\begin{figure}[t]
  \centering
  \includegraphics[width=\columnwidth]{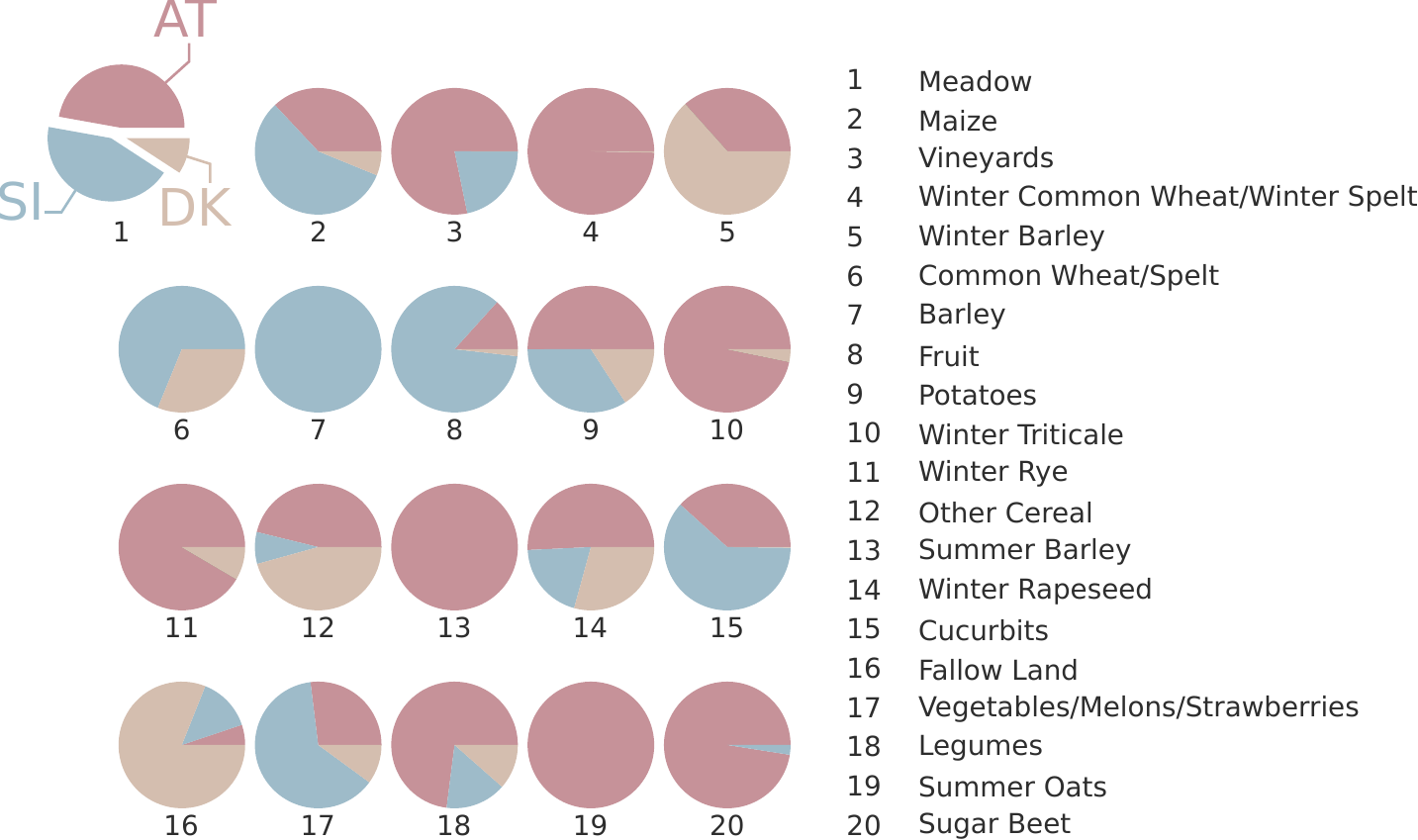}
  \caption{
    The differences between the three countries regarding the most grown crops becomes overly visible when analysing the relative amount of parcels that each country contributes to the respective class.
    Here we show these distributions for the top-20 crop types (\cf \cref{fig:dataset_stats}).
    Clearly, not all regions contribute equally to each crop class and some countries already impose a coarser granularity.
    While \emph{Austria}, for instance, includes the season with \class{Common Wheat/Spelt} (\cf class 4), \emph{Slovenia} and \emph{Denmark} (\cf class 6) omit this information.
    It might, therefore, be recommended to decrease the level of granularity when dealing with the entire dataset at once.
    In other cases and when just working with a subset of the data, a distinction between summer and winter crops might be of interest, influencing the descision to not aggregate crop classes when designing the dataset.
  }
  \label{fig:pies}
\end{figure}
}

%% file: tables.tex
\newcolumntype{x}{l}
\newcolumntype{X}{>{\scriptsize}l}
\newcolumntype{y}{r}
\newcolumntype{Y}{>{\scriptsize}r}
\newcolumntype{v}[1]{>{\raggedright\arraybackslash\hspace{0pt}}p{#1}}
\newcolumntype{V}[1]{>{\scriptsize\raggedright\arraybackslash\hspace{0pt}}p{#1}}

\newcommand{\tabAvailability}{%
\begin{table}
  \caption{Member states of the EU sorted by the availability of their agricultural reference data, as of April 2021.}
  \label{tab:countries}
  \centering
  \begin{tabular}{@{}Xv{.75\linewidth}@{}}
    \toprule
    \multicolumn{1}{@{}X}{Availability} & \multicolumn{1}{X@{}}{Countries}\\
    \cmidrule(r){1-1} \cmidrule(l){2-2}
    available &
    Austria, Belgium, Croatia, Denmark, Estonia, France, Latvia, Lithuania, Netherlands, Portugal, Sweden, Slovakia, Slovenia \\
    \addlinespace
    planned &
    Finland, Romania, Spain \\
    \addlinespace
    not planned &
    Czechia, Germany, Greece, Hungary, Ireland, Italy, Luxembourg, Malta, Poland \\
    \addlinespace
    request pending &
    Cyprus, Bulgaria \\
    \bottomrule
  \end{tabular}
\end{table}
}

\newcommand{\tabCountryTiles}{%
  \begin{table}[t]
    \caption{%
      Listed below are the chosen countries and the corresponding \sentinel tile name covering the demo region, together with the number of usable parcels and record year. 
      \num{150000} fields without meaningful annotations are omitted.
    }
    \label{tab:sentineltilemnames}
    \centering
    \begin{tabular}{@{}xxyy@{}}
      \toprule
      \multicolumn{1}{@{}X}{Country}   & 
      \multicolumn{1}{X}{\sentinel Tile Name} & 
      \multicolumn{1}{X}{Number of Parcels} & 
      \multicolumn{1}{X@{}}{Year}  \\
      \cmidrule(r){1-1} \cmidrule(lr){2-2} \cmidrule(lr){3-3} \cmidrule(l){4-4}
      Austria     	& \tilename{T33UWP} & \num{396600} & 2020\\
      Denmark     	& \tilename{T32VNH} & \num{98565} & 2019\\
      Slovenia     	& \tilename{T33TWM} & \num{310236} & 2020\\
      \bottomrule
    \end{tabular}
  \end{table}
}

%% file: notation.tex
\usepackage[detect-all, binary-units]{siunitx}
\usepackage{xspace}

\newcommand{\filename}[1]{\texttt{#1}\xspace}
\newcommand{\tilename}[1]{\texttt{#1}\xspace}

\newcommand{\programminglanguage}[1]{\textsc{#1}\xspace}

\newcommand{\geojson}{\programminglanguage{GeoJSON}}

\newcommand{\brand}[1]{\textsc{#1}\xspace}
\newcommand{\satellitename}[1]{\brand{#1}}
\newcommand{\sentinel}{\satellitename{Sentinel-2}}
\newcommand{\eurocrops}{\brand{EuroCrops}}

\newcommand{\class}[1]{\textsl{#1}}

\newcommand{\github}{\brand{GitHub}}

\newcommandx{\HCAT}[6][3=,4=,5=,6=]{%
  {\ttfamily%
  #1#2%
  \notblank{#3}{--#3}{}%
  \notblank{#4}{--#4}{}%
  \notblank{#5}{--#5}{}%
  \notblank{#6}{--#6}{}%
  }%
}



%% file: images/dataset_statistics_slovenia.pgf
\pgfplotstablesort[
  col sep=comma, 
  sort cmp=float >,
  sort key={Absolute Occurrence}, 
]
{\datatablesorted}%
{./data/all_top20.txt}
%
\begin{tikzpicture}
  \begin{axis}[
    width=.35\linewidth,
    height=4cm,
    crop bar,
    ymode = log, log origin=infty, log ticks with fixed point, 
    ymax=1, ymin=.5e-3,
    xtick=data,
    xticklabels from table={\datatablesorted}{Crop Name},
    xlabel=\empty, xlabel style={overlay},
    ylabel=\empty, ylabel style={overlay},
]
    \addplot[freq bar, fill=pieblue] table [x expr=\coordindex, y={Relative Occurrence}]{\datatablesorted};
    \addlegendentry{Relative Occurrence}
  \end{axis}
\end{tikzpicture}%

%% file: images/parcels.pgf
\tikzset{
  cascade/.style={draw,thin,minimum height=7mm,minimum width=7mm},
}%
\newcommand\randmin{}%
\newcommand\randmax{}%
\newcommand\randmultof{}%
\newcommand\setrand[4]%
  {\def\randmin{#1}%
   \def\randmax{#2}%
   \def\randmultof{#3}%
   \pgfmathsetseed{#4}%
  }%
\newcommand\nextrand
  {\pgfmathparse{int(int((rnd*(\randmax-\randmin+1)+\randmin)/\randmultof)*\randmultof)}%
   \xdef\thisrand{\pgfmathresult}%
  }%
\newcommand{\stack}[1]{%
  \setrand{40}{70}{10}{42}
	\foreach \x in {1,...,1}{
	  \nextrand
	  \node[cascade, fill=black!\thisrand!white] (#1\x) at (-\x mm, -\x mm) {};
	};
	\node[below left=6mm and 6mm of #11.center, rotate=80, anchor=center, inner sep=0pt, font=\bfseries\large] {$\ddots$};
	
	\foreach \x in {12,...,14}{
	  \nextrand
	  \node[cascade, fill=black!\thisrand!white] (#1\x) at (-\x mm, -\x mm) {};
	};
}%
\begin{tikzpicture}

\stack{a}
\node[fit=(a1)(a14),inner sep=0pt] (bb_a) {};

\begin{scope}[xshift=1.5cm]
  \stack{b}
\end{scope}
\node[fit=(b1)(b14),inner sep=0pt] (bb_b) {};

\begin{scope}[xshift=4cm]
  \stack{c}
\end{scope}
\node[fit=(c1)(c14),inner sep=0pt] (bb_c) {};

\node[font=\large\bfseries] at ($(b14.east)!.5!(c14.west)$) {\ldots};

\node[fit=(bb_a)(bb_b)(bb_c)] (bb) {};
\draw[|-|] ([yshift=-1mm]bb.south west) -- node[below] {$T$ timesteps} ([yshift=-1mm]bb.south east);

\draw[|-|] ([xshift=-2mm]bb_a.north west) -- node[left] {parcel $x$} ([xshift=-2mm]bb_a.south west);

\draw[|-|] ([xshift=3mm]c14.south east) -- node[right] {$C$ bands} ([xshift=3mm]c1.south east);

\end{tikzpicture}%